\documentclass[aps,pra,english,twocolumn,showpacs]{revtex4-1}
\usepackage[colorlinks=true, citecolor=blue, urlcolor=blue ]{hyperref}
\def\be{\begin{equation}}
\def\ee{\end{equation}}
\def\bea{\begin{eqnarray}}          
\def\eea{\end{eqnarray}}
\def\bi{\begin{itemize}}
\def\ei{\end{itemize}}
\def\nb{\nonumber}
\usepackage{xcolor}
\usepackage{graphicx}
\usepackage{float}
\usepackage{soul}
\usepackage{amsmath,amstext,amssymb,babel,bm,color,times}
\DeclareMathOperator{\Tr}{Tr}

\begin{document}

\title{ 
   Framework of dynamical transitions from long-range to short-range quantum systems
}
\author{Leela Ganesh Chandra Lakkaraju\(^1\), Srijon Ghosh\(^1\), Debasis Sadhukhan\(^2\), Aditi Sen(De)\(^1\)}

\affiliation{\(^1\) Harish-Chandra Research Institute,  A CI of Homi Bhabha National Institute, Chhatnag Road, Jhunsi, Prayagraj - 211019, India}

\affiliation{\(^2\) School of Informatics, University of Edinburgh,
10 Crichton Street, Edinburgh EH8 9AB, Scotland, UK}

\begin{abstract}
A quantum many-body system undergoes phase transitions of distinct species with variations of local and global parameters.
We propose a framework in which a dynamical quantity can change its behavior for quenches across global (coarse-grained criterion) or local system parameters (fine-grained criterion), revealing the global transition points.  We illustrate our technique by employing the long-range extended Ising model in the presence of a transverse magnetic field. We report that by distinguishing between algebraic and exponential scaling of the total correlation in the steady state, one can identify the first transition point that conventional indicators such as the rate function fail to detect. To determine the second one, we exploit the traditional local quenches. During quenches with and without crossing the critical points along the local parameter, total correlation follows either the same or different scaling laws depending on its global phase. 
\end{abstract}

\maketitle

\section{Introduction}
\label{sec:intro} 
Long-range (LR) systems are often intriguing avenues to test the validity of the hypotheses that are otherwise well-understood for prototypical short-range (SR) systems. 
The astounding progress in recent experiments, particularly within cold-atomic platforms, has now made it feasible to engineer and fine-tune LR systems with exceptional precision \cite{RevModPhys.80.885, *Bloch2012, RevModPhys.82.2313, Weber2010, Yao2012, Dolde2013, PhysRevLett.108.210401, PhysRevLett.108.215301,  Schau2012, Yan2013, Firstenberg2013, Douglas2015, Drfler2019, Pagano2019, Tao2020, Islam2011, Britton2012, Islam2013, PhysRevLett.125.133602, RevModPhys.93.025001}. As a result verifying these hypotheses in laboratories has become a tangible reality. Not only the ground states,
these recent advancements are exquisite enough to probe the far-from-equilibrium time dynamics of these LR systems in a controlled manner \cite{Richerme2014, Jurcevic2014, Bernien2017, doi:10.1146/annurev-conmatphys-031214-014548}.  
Consequently, over the past few decades, theoretical and experimental research has generated a great deal of interest in non-equilibrium quantum phenomena, which has led to the discovery of some fascinating non-equilibrium physics including Kibble-Zurek mechanism~\cite{K-a, *K-b, *K-c, Z-a, *Z-b, *Z-c, PhysRevLett.95.245701, Keesling2019, King2022, PhysRevB.106.L041109}, sonic horizons of correlation spreading~\cite{PhysRevB.93.075134, PhysRevB.98.024302, PhysRevX.8.021069, PhysRevX.8.021070, SonicHorizon, PhysRevX.10.031010, PhysRevX.10.031009, *PhysRevX.13.029901, PhysRevA.104.062420}, the breaking of ergodicity~\cite{Turner2018, Serbyn2021, PhysRevA.87.052318,Prabhu2013,PhysRevA.94.042310,PhysRevA.97.032103, PhysRevB.99.064422},  many-body localization~\cite{doi:10.1146/annurev-conmatphys-031214-014726, RevModPhys.91.021001, Schreiber2015, Smith2016, PhysRevX.7.041047},  discrete time-crystals~\cite{doi:10.1146/annurev-conmatphys-031119-050658, PhysRevA.97.053621, PhysRevA.98.023612, PhysRevLett.126.020602, Kyprianidis2021}, and many more. 

A commonly considered out-of-equilibrium dynamics, commonly referred to as {\it sudden quench}
is usually described by an instantaneous change in the underlying Hamiltonian, say, $H^{in}\to H^f$,  leading to the time-evolved state $|{\Psi}(t)\rangle=e^{-iH^f t}|{\Psi}^{in}\rangle$, where $|{\Psi}^{in}\rangle$ is the ground state of $H^{in}$ just before the quench. 
Physics at transient timescales is of particular interest primarily because their experimental observations are within the reach of current techniques, while the properties of the steady state are crucial from a statistical perspective.  
Although quantum phase transitions are well understood, their implications in the non-equilibrium regime especially their universal scaling laws and universality classes, remain relatively uncharted. 
In recent years, there have been numerous attempts to extend the concept of equilibrium quantum phase transition (EQPT) to the out-of-equilibrium domain, collectively referred to as dynamical quantum phase transitions (DQPT) \cite{Heyl2018, Marino_2022}. 
 They are examined via a suitable order parameter (which we call order-parameter-based DQPT, oDQPT),  exhibiting some behavioral changes in the wake of the oDQPT \cite{PhysRevA.72.052319, PhysRevLett.100.175702, PhysRevLett.105.220401, PhysRevLett.106.227203, ShekharDhar2012, PhysRevB.88.201110, Dhar2014, PhysRevB.91.205136, PhysRevA.93.042322, PhysRevB.94.214301, PhysRevB.95.024302}.
 On the other hand, in recent times \cite{paolo_dqpt_1_2006,PhysRevLett.110.135704}, the non-analyticities in the rate function $\ln |\langle\Psi^{in}|\Psi(t)\rangle|^2$, 
 akin to the Fisher zeros~\cite{Fisherlectures} in the complex temperature plane for equilibrium transitions, are used to identify DQPTs, referred to as rate-function based DQPT (rDQPT) for which vanishing return probabilities ensure the transition. However, there is no associated Landau-type order parameter to be evaluated \cite{LandauPhy4, *1965}. 
 Since the inception of rDQPT \cite{PhysRevLett.110.135704}, there have been numerous endeavors to study the same across a diverse array of many-body systems 
 \cite{PhysRevB.87.195104, https://doi.org/10.48550/arxiv.1308.0277, Pozsgay2013, PhysRevLett.113.265702, PhysRevB.89.054301, PhysRevB.90.125106, PhysRevB.91.155127, PhysRevB.92.104306, PhysRevLett.115.140602, PhysRevB.92.075114, PhysRevB.93.144306, Zhang2016, PhysRevB.94.165135, PhysRevB.95.060504, PhysRevB.96.134427, PhysRevB.96.180303, PhysRevB.97.134306, PhysRevB.99.054302, PhysRevB.102.144306, PhysRevResearch.2.033111, PhysRevResearch.4.013250, Hauke2021, PhysRevB.101.224304, Porta2020, PhysRevB.107.094304, PhysRevB.107.094307, PhysRevB.105.165149} including some experimental validations \cite{PhysRevLett.114.010601, Flschner2017, PhysRevLett.119.080501, Zhang2017, PhysRevLett.124.043001, PhysRevLett.124.250601}. 
 Contrary to its initial assertion, subsequent studies  \cite{PhysRevB.89.161105, Jafari2019, PhysRevB.89.125120} have shown that rDQPTs may not always be linked to EQPTs. There have been 
multiple attempts to explain rDQPT from different perspectives including symmetry breaking~\cite{PhysRevLett.113.205701, PhysRevLett.123.040601, PhysRevLett.129.260407, PhysRevLett.130.100402}, spreading of entanglement \cite{PhysRevLett.122.250601, PhysRevLett.126.040602}, 
etc. However, a definitive consensus remains elusive.

 \begin{figure}[t]
    \centering
    \includegraphics[width=\linewidth]{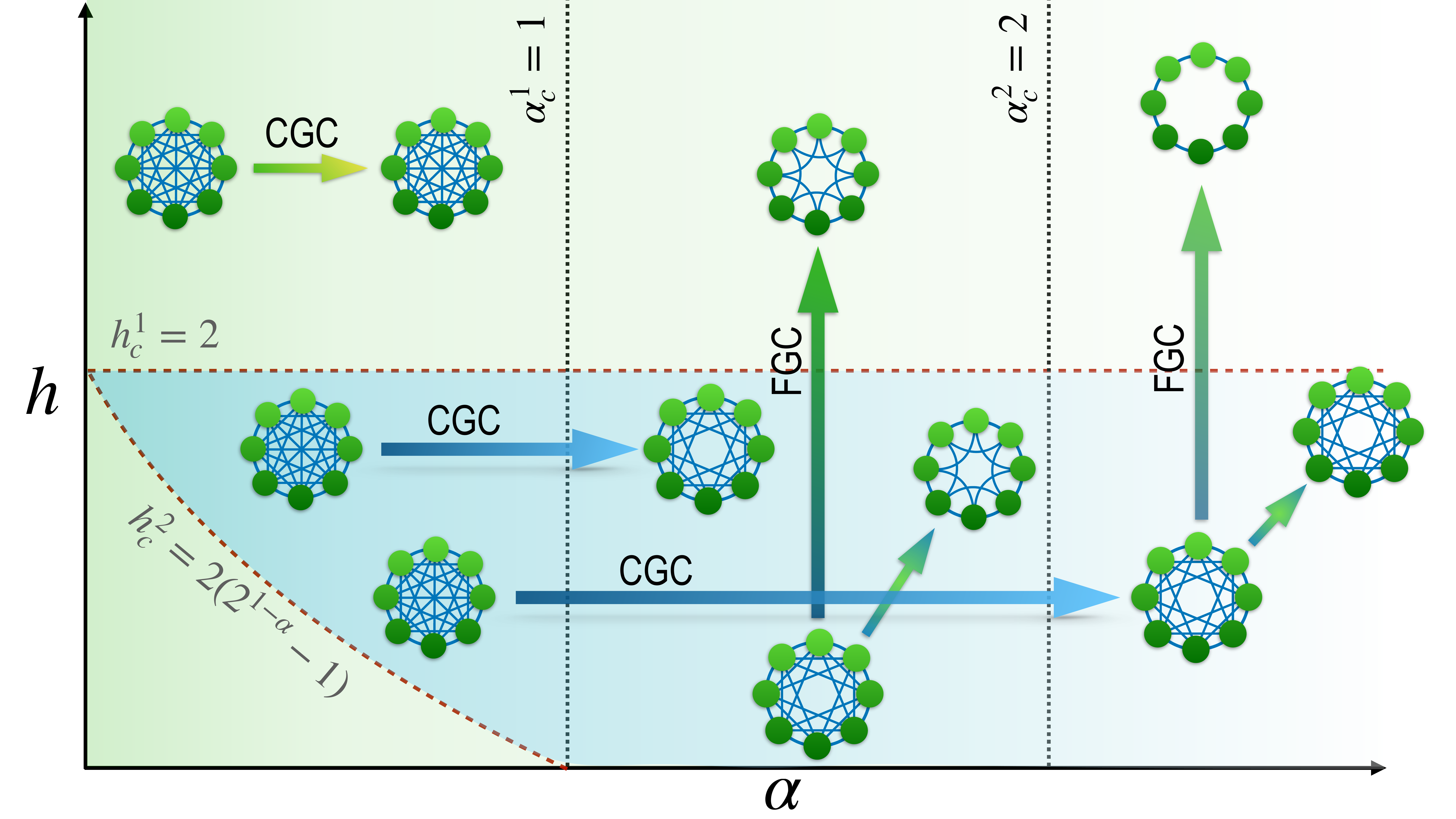}
   \caption{Potential of total correlation (TC) for detecting global phases after LR to SR quench. The system possesses a rich phase diagram containing transitions along a local parameter, \(h\), and also along the global tuning, \(\alpha\). 
The scaling of TC is used as our decisive criterion. An all-to-all connectivity accounts for an algebraic scaling of TC throughout the lattice chain where as fewer connectivities represent an exponential decay of TC vanishing at finite distances (refer Fig.\ \ref{fig:mutual_alpha_quench} and \ref{fig:mutual_alpha_quench1}).  
The coarse-grained criteria (CGC) deals with global quenches across \(\alpha\) while the fine-grained criteria (FGC) concerns local quenches across $h$. The joint criteria are capable of identifying both the global transitions at $\alpha_c^{\{1,2\}}$. }
   \label{fig:schematic}
\end{figure}

In this paper, we present a general framework of DQPT that relies on total correlations (TC) \footnote{In the literature, mutual information sometimes wrongly referred to quantum correlation or entanglement. we use the word 'total correlation' to avoid any confusion and to stress the fact that it includes both the classical and quantum correlation.} quantified via the quantum mutual information \cite{olliver2001, Henderson2001, groisman2005}  between $i^\text{th}$ and $(i+R)^\text{th}$ lattice site, given by $I_{R} = S(\rho_i)+S(\rho_{i+R})-S(\rho_{i,i+R})$ with $S(*) = \Tr(*\log *)$ being the von-Neumann entropy and $\rho_i$ and $\rho_{i+R}$ being the local reduced density matrices of $\rho_{i,i+R}$. 
We demonstrate that the scaling of the same can serve the purpose of detecting DQPTs from LR to SR systems. 
We label a quench {\it local} when it pertains exclusively to local parameters within a Hamiltonian while the term {\it global} quench is used for significant overarching modifications in the system such as transitioning from an all-to-all interacting system to a nearest-neighbor one which remarkably
are already within the purview of ongoing experiments \cite{Islam2011, Britton2012, Islam2013, RevModPhys.93.025001, PhysRevLett.125.133602, Richerme2014, Jurcevic2014, Rajabi2019, PhysRevLett.129.220501, Katz2023, PRXQuantum.4.030311}.
We are unaware of any well-established universality classes for such global transitions and to the best of our knowledge, quenches in this global sense have never been explored before.
This study delves into such atypical global quenches and seek for suitable order parameters in the steady state that can manifest the global transitions.
In particular, we ask: {\it When a system is subject to a global quench, does the steady state carry the signature of the LR to SR transition}? 

It turns out that traditional identifiers such as the rate function fails to identify such global transition in a ubiquitous manner. Here we show that our comprehensive framework, comprising fine-grained and coarse-grained criteria, are adequate to single out global quenches involving transition points (see Fig.\ \ref{fig:schematic}). 
In particular, we focus on quenches from LR to SR Hamiltonians which do not involve traditional EQPTs associated with the gap-closing of the energy. Instead, the LR to SR quenches involve transitions between three regimes, namely non-local, quasi-local, and local, which are characterized by the information propagation speed across the system. While the rate function alone cannot detect these transitions, TC can discern the transition between non-local and local (or quasi-local) regimes via the coarse-grained criterion.  
To differentiate between quasi-local and local regimes, we invoke an additional fine-grained criterion that involves quenches across the local parameter. We demonstrate that the joint criteria are exquisite enough to differentiate between the dynamics to different global regimes.   

\section{Framework}
\label{sec:framework}
Let us consider a many-body quantum system consisting of two noncommuting parts, $H_\text{local}$ and $H_\text{int}$, given by
\begin{equation}
    H = H_\text{int} + H_\text{local} \equiv H_\text{int}({\alpha}) + \sum_i H_i({h}).
    \label{GeneralH}
\end{equation}
Here  $H_\text{local}$  is constituted by a set of 
local operators $\{H_i (h)\}$ while 
$H_\text{int} (\alpha)$ 
represents global operators  
that influence the global properties of the entire system. 
Typically, $H_i$ acts on individual (or few) lattice sites while interaction part $H_\text{int} (\alpha)$ being global cannot be probed individually for a few sites. 
As a prototypical example, the local parameter $h$ could be the external magnetic field while the global parameter $\alpha$ would correspond to the strength of the LR interactions. 
At equilibrium, the system possesses multiple transitions within the $(h,\alpha)$-plane (Fig.\ \ref{fig:schematic}).
While criticalities across ${h}$ are associated with the energy gap-closing, transitions across $\alpha$ are related to the information propagation speed (see Appendix \ref{phases}).  
The presence of distinct order parameters for transitions across $h$ and $\alpha$ possibly suggests that dissimilar dynamical quantities would be necessary to identify dynamical phases across these transitions, and this is indeed what we establish in this work.

After preparing the initial state as the ground state of a Hamiltonian $H^{in} = H(h,\alpha^i)$, a sudden quench is performed to $H^{f} = H(h,\alpha^f)$. 
Similar to oDQPT, we look for a suitable order parameter in the steady state that would carry the signature of the $\alpha$-transitions. 
Choosing scaling of TC (for definition, see Appendix \ref{mutual_info}) as the touchstone, we propose the following two criteria as the identifier of LR to SR transition in dynamics: 

\emph{Course-grained criteria (CGC)} --
Initiating a global quench from the non-local regime, if TC between sites $i$ and $i+R$ in the steady state is found to be extended across the entire lattice with an algebraic decay, there are no global transitions involved in the quench. However, if the same vanishes at an exponential pace, the quench must have included at least one transition point in the global quench. 

\emph{Fine-grained criteria (FGC)} --
When the system, prepared at the global phase corresponding to $\alpha^f$, is subjected to a local quench across $h$, the scaling of TC at steady state is sensitive to EQPTs only in the local regimes, but not in the quasi-local regime.


\section{Model}
\label{sec:model}
At this point, let us specify a Hamiltonian which contains both the local and global constituents. We choose the periodic transverse LR extended-Ising chain, 
\bea
{ H(h,\alpha)}
=
\sum_{i=1}^N
\left[ \frac{h}{ 2}\sigma^z_i +
 \sum_{R=1}^{N/2} 
J_R(\alpha)~{\cal S}_{i,R}\right],
\label{Hamil}
\eea
with \({\cal S}_{i,R}=\sigma^x_i\prod_{j=i+1}^{i+R-1}\sigma^z_{j}\sigma^x_{i+R}\) and \(J_R(\alpha)={R^{-\alpha}}/A\)
as our work-horse in this study. 
Here $\sigma_i^z$ acts on $i^\text{th}$ site while the operator ${\cal S}_{i,R}$ encompasses multiple sites starting from $i^\text{th}$ to $(i+R)^\text{th}$ lattice and $A$ is the Kac normalization.
The model possesses a rich phase diagram which include quantum critical points at \(h_c^{\{1,2\}}= \{2, 2(2^{1-\alpha}-1)\}\) as well as two global transition points at \(\alpha_c^{\{1,2\}} = \{1,2\}\) separating among the non-local ($\alpha<1$), quasi-local ($1<\alpha<2$) and local ($\alpha>2$) phases
(see Appendix \ref{hamiltonian}, \ref{diagonalization} and \ref{phases} for a brief description of the Kac normalization, diagonalization procedure and phase diagram of the considered system respectively). 

 \begin{figure}[t]
    \centering
    \includegraphics[width=\linewidth]{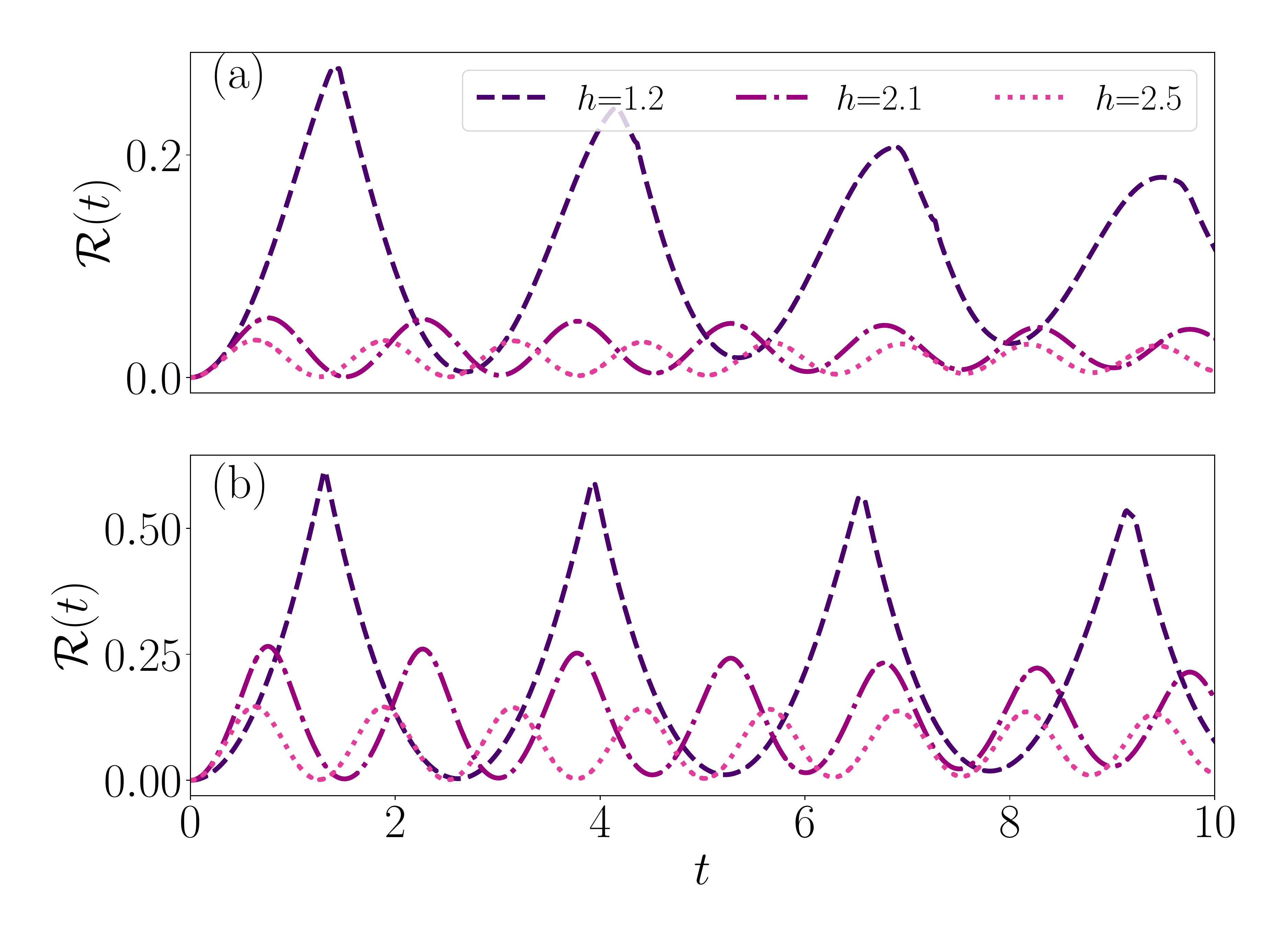}
    \caption{Rate function $\mathcal{R}(t)$ as function of time $t$ after LR to SR quench. The system is initially prepared in the non-local regime with $\alpha^{i} = 0.5$ and then subject to a global quench to (a) quasi-local regime ($\alpha^{f} = 1.5$) and (b) local regime ($\alpha^f = 3.0$). $N = 512$.}
    \label{fig:alpha_quench_rate_function}
\end{figure}

We are interested in the steady state TC between two distant sites, $i$ and $i+R$, which requires the reduced density matrix of the corresponding sites, $\rho_{i,i+R} = \Tr_{\overline{i,i+R}}[\rho^\infty]$ where $\rho^\infty= \lim_{t\to\infty} |\Psi(t)\rangle\langle \Psi(t)|$ 
is the steady state reached after waiting for a sufficiently long time following the quench and the partial trace is taken over all lattice sites except $i$ and $i+R$. Since the Hamiltonian possesses translational invariance due to periodic boundary condition $(\sigma^\alpha_{N+R} = \sigma^\alpha_R)$, only the distance $R$ matters (without loss of generality, we can omit index $i$) and hence we use $\rho_R(t)$ for any two-site density matrix which is distance $R$ apart. The general form of $\rho_R(t)$ can be expressed through Pauli operator basis as  ${\rho}_{R} (t)=\big[\mathbb{I}_4 +m^z (t) (\sigma^z \otimes \mathbb{I}_2+\mathbb{I}_2 \otimes \sigma^z)+ \sum_{\alpha,\beta}^{x, y, z}\mathcal{C}^{\alpha \beta}_{R} (t)\sigma^\alpha\otimes\sigma^\beta \big]/4$ where $m^z$ is the transverse magnetization and $C_R^{\alpha \beta}=\text{tr}(\sigma^\alpha\sigma^\beta\rho_R)$ are the two site classical correlators for $\alpha,\beta = x,y,z$. 
 As $m^x, m^y, C^{xz}_R$,  $C^{zx}_R$, $C^{yz}_R$ and $C^{zy}_R$ contain odd number of fermionic operators, they all vanish at all times (by Wick's theorem). 
Hence, the only nonvanishing correlations, whose time-dynamics we need, are \(m^z(t), C^{\alpha \alpha}_R (t)$ \((\alpha=x, y, z)\),  \(C^{xy}_R (t)\) and  \(C^{yx}_R (t)\).
For eigenstates of Eq.\ (\ref{Hamil}), such as the ground state or the thermal state, the two-site correlation functions take the form of $R\times R$ Toeplitz determinant. However, since our focus lies on the time-evolved state, which typically do not correspond to eigenstates of the final Hamiltonian, the use of Toeplitz determinants alone is insufficient (see \cite{EHRHARDT1997229} for an extended discussion). As Szeg\"o theorem \cite{szego1952certain} becomes inapplicable in this case, we must directly deal with the Pfaffians \cite{Caianiello1952} to evaluate the correlators. If we ignore the signs of the correlation functions, they can again be computed from the square roots of the block $2R\times 2R$ Toeplitz determinants of the complete Pfaffian. However, in our case, the sign cannot be ignored, hence we resort to the numerical algorithm introduced in Ref.\ \cite{num_pfaff_wimmer_2012} for the evaluation of the Pfaffians (see Appendix \ref{pfaff_corre} for the full prescription). After constructing $\rho_R(t)$ from the classical correlators, we compute $I_R$ with varying distance $R \in \{1, \ldots, \frac{N}{2}\}$ in the steady state. 

\begin{figure}[t]
    \centering
     \includegraphics[width=\linewidth]{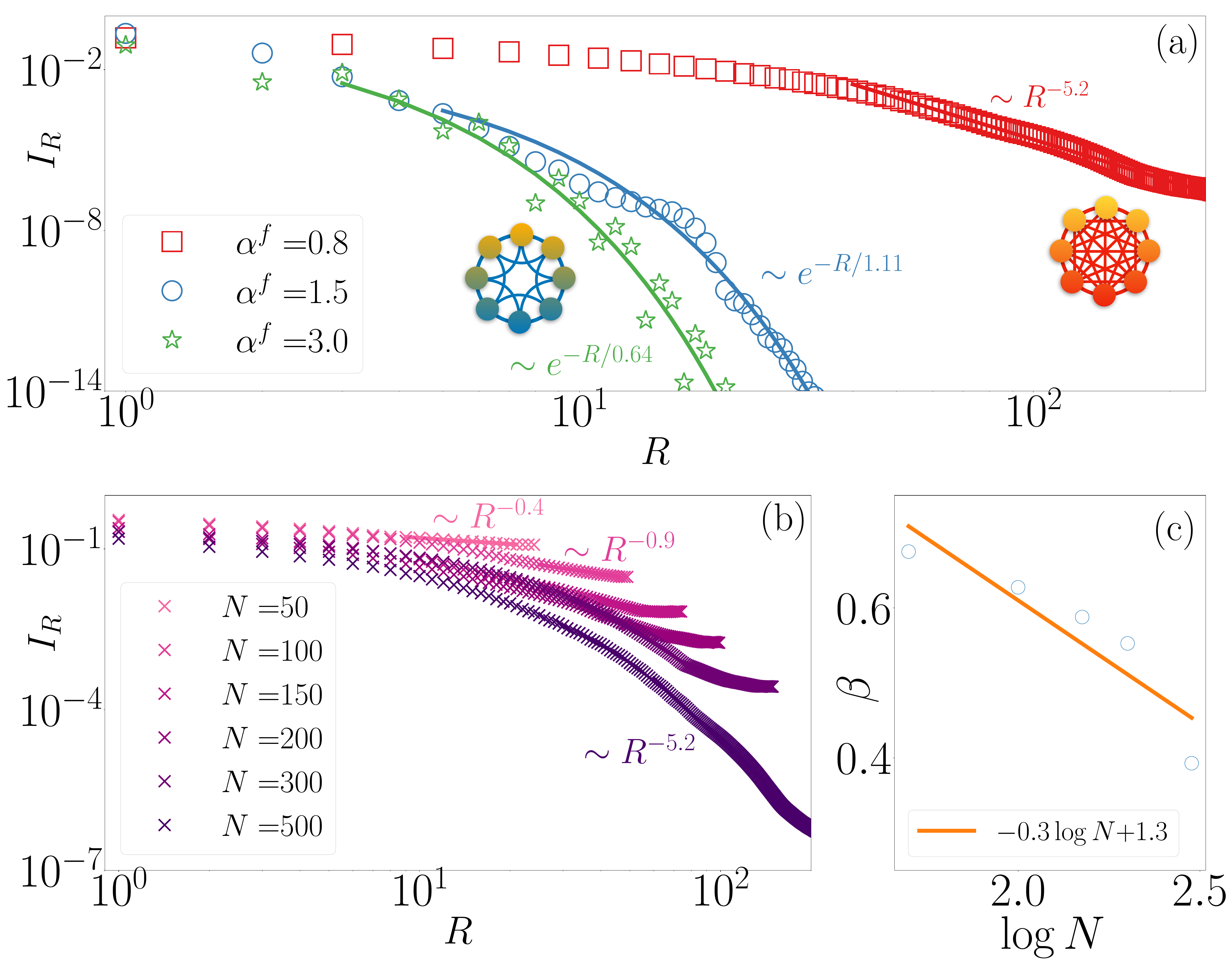}
    \caption{ Scaling of steady state TC ($I_R$) with distance $R$ in the ordered phase ($h=0.5$). The time evolved state at $t=200$ is considered as the steady state as $I_R$ saturates well before. (a) The quench protocol is same as Fig. \ref{fig:alpha_quench_rate_function} with $\alpha^i=0.5$ and $\alpha^f=\{0.8,1.5,3\}$. For quenches within non-local regime (red squares), $I_R$ is non-vanishing throughout the lattice with an algebraic tail $I_R \sim R^{-\eta}$, however, a quench to the quasi-local (blue circles) or local (green stars) regime results in exponential scaling $I_R\sim e^{-R/\xi}$. The fitted functions are the respective scalings shown in solid lines of corresponding color. (b) Algebraic scaling of the $\alpha^f=0.8$ quench for different $N=\{50, \ldots, 500\}$. The tails are best fitted with $R^{-\eta_N}$ for each $N$. (c) Finite-size scaling analysis of $\eta_N$. The vertical axis represents $\beta=\log|\eta_N-\eta_\infty|$. Considering $N=500$ as the thermodynamic limit, a convergence towards $\eta_\infty$ is shown as  $N^{-0.3}$ indicating sustainability of the algebraic scaling in the thermodynamic limit.}
    \label{fig:mutual_alpha_quench}
\end{figure}

\section{Rate Function} 
\label{sec:ratefucntion}
Let us first find out how well the traditional identifier such as the rate function performs to detect the LR to SR transition.  
The initial state is prepared corresponding to the LR Hamiltonian in the the non-local regime ($\alpha^i<\alpha_c^1=1$) and then suddenly quenched to either quasi-local or local regime. 
The local parameter $h$ is fixed to a predetermined value such that the local phase is fixed either to the ordered phase ($h_c^2>h>h_c^1$) or to the disordered phase ($h>h_c^1,h<h_c^2$). Drawing analogy from DQPT across the local parameter, one would expect that the rate function would show non-analyticity if the global quench involves any transition points. It turns out the rate function fails to do so in a universal manner, i.e., irrespective of its local phases. In particular, in the ordered phase (say, $h=1.2$), the rate function shows non-analyticities for global quenches across transition points, although in the disordered phase ($h=\{2.1,2.5\}$), it is always smooth, thereby not carrying the signatures of the global transition as shown in Fig.\ \ref{fig:alpha_quench_rate_function} (see Appendix \ref{rate_function} and \ref{h_quench} for more details).

 Since the $\alpha$-transitions are not linked with the gap-closing of the energy, which is the primary foundation behind  formulating rDQPT \cite{PhysRevLett.110.135704}, we argue that the rate function is not a good candidate for identifying the $\alpha$-transitions.
We have to seek a quantifier that manifests the information propagation speed across the system and can serve as a decisive factor in distinguishing different global transitions. It turns out that the scaling of TC can indeed fulfill this role.

\section{TC-based criteria}

Let us now demonstrate that the above-defined CGC and the FGC with the scaling of TC as the order parameter can be used to achieve our objective in a ubiquitous manner including the disordered phase where the rate function fails.
See Appendix \ref{mutual_info} for computations of TC in terms of correlation functions.
\begin{figure}[t]
    \centering
    \includegraphics[width=\linewidth]{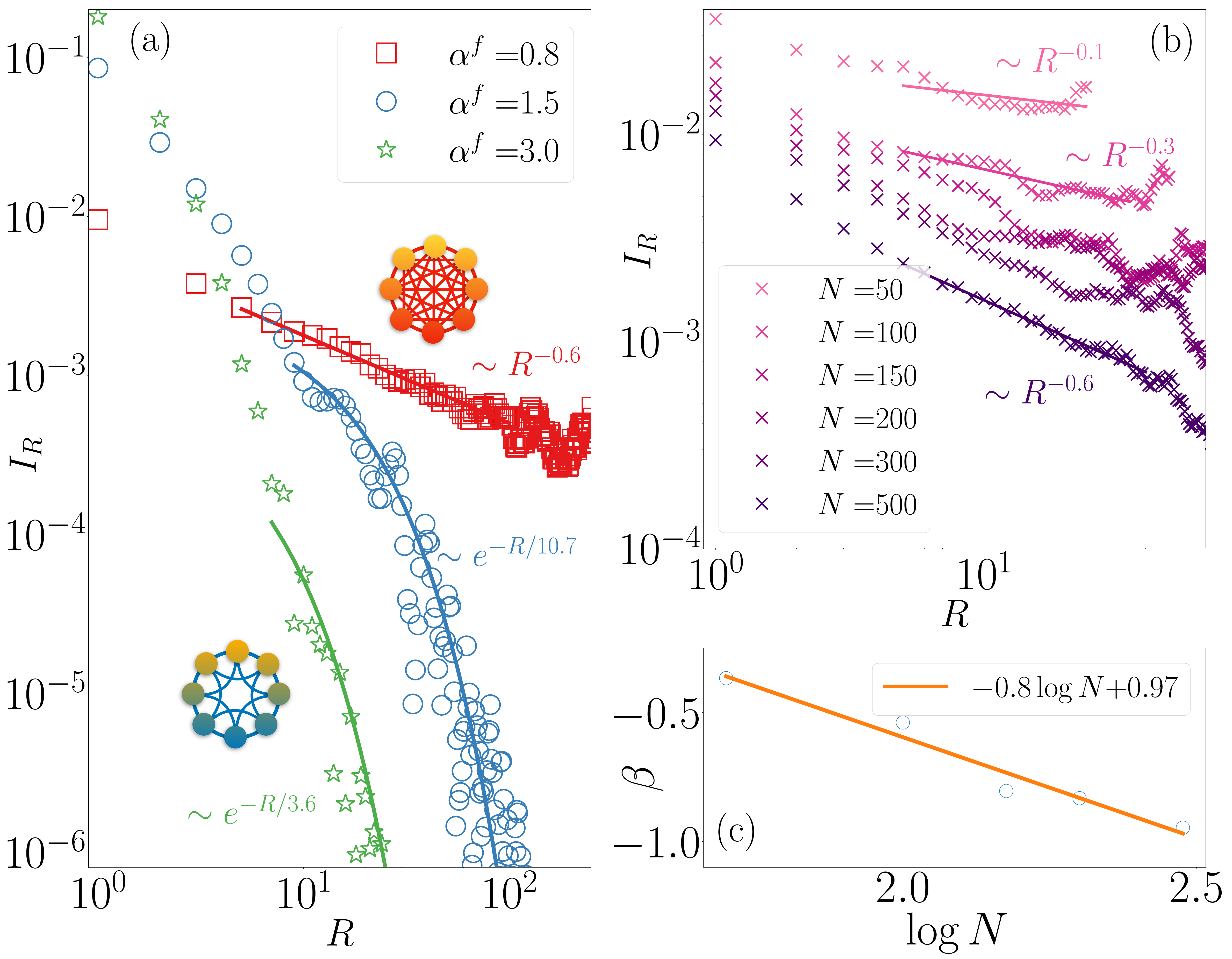}
    \caption{Scaling of steady state TC in the disordered phase ($h=2.5$). All other quantities, except $h$, are same as Fig.\ \ref{fig:mutual_alpha_quench}.} 
    \label{fig:mutual_alpha_quench1}
\end{figure}

\subsection{Course-grained criteria (CGC)} 
In Fig.\ \ref{fig:mutual_alpha_quench}  (\ref{fig:mutual_alpha_quench1}), the steady state TC is plotted as function of distance $R$ with the local parameter $h$ fixed at the ordered (disordered) phase. When the global quench is within the non-local regime, TC of the steady state follows an algebraic scaling irrespective of the local parameter $h$. If the quench involves {\it at least} one $\alpha$-transition point, TC vanishes quickly with an exponential decay as depicted in Figs \ref{fig:mutual_alpha_quench}(a) and \ref{fig:mutual_alpha_quench1}(a). Mathematically, the scaling laws read as
\begin{eqnarray}
  I_R^{\alpha^i\to \alpha^f} \sim \left \{
    \begin{array}{cc} 
        R^{-\eta}   \quad &\implies \quad \alpha^{f} < \alpha_{c}^{1},  \\
        e^{-R/\xi} \quad &\implies \quad \alpha^{f} > \alpha_{c}^{1}. 
    \end{array} \right.
    \label{scale_c2}
\end{eqnarray}
In order to show that the algebraic scaling is retained in the thermodynamic limit, finite-size scaling analyses of the exponent $\eta$  are shown in panels $(b)-(c)$ of Figs.\ \ref{fig:mutual_alpha_quench} and \ref{fig:mutual_alpha_quench1}. We notice, that for larger systems, TC closely resembles the pattern obtained for $N=500$. Hence, we can safely assume that it can serve the purpose of infinite spin chains. Our analysis shows that $\eta_N$ converges to $\eta_\infty$ as $N^{-0.3}$ and $N^{-0.8}$ respectively in the ordered and disordered phases.

\subsection{Fine-grained criteria (FGC)} 
Although CGC can ascertain quenches across $\alpha^1_c$, it cannot conclusively tell whether the quench encompasses only a single $\alpha$-transition or multiple. In other words, when TC is exponentially decaying, CGC concludes $\alpha^f>\alpha_c^1$, but cannot pinpoint whether the global phase corresponding to $\alpha^f$ is local or quasi-local. We now illustrate that the TC scaling of steady state after a local quench is sensitive to the EQPTs across $h$ in the local regime ($\alpha>2$) but not in the quasi-local regime ($1<\alpha<2$). We exploit the same to differentiate between the quasi-local and local regimes.

\begin{figure}[t]
    \centering
    \includegraphics[width=\linewidth]{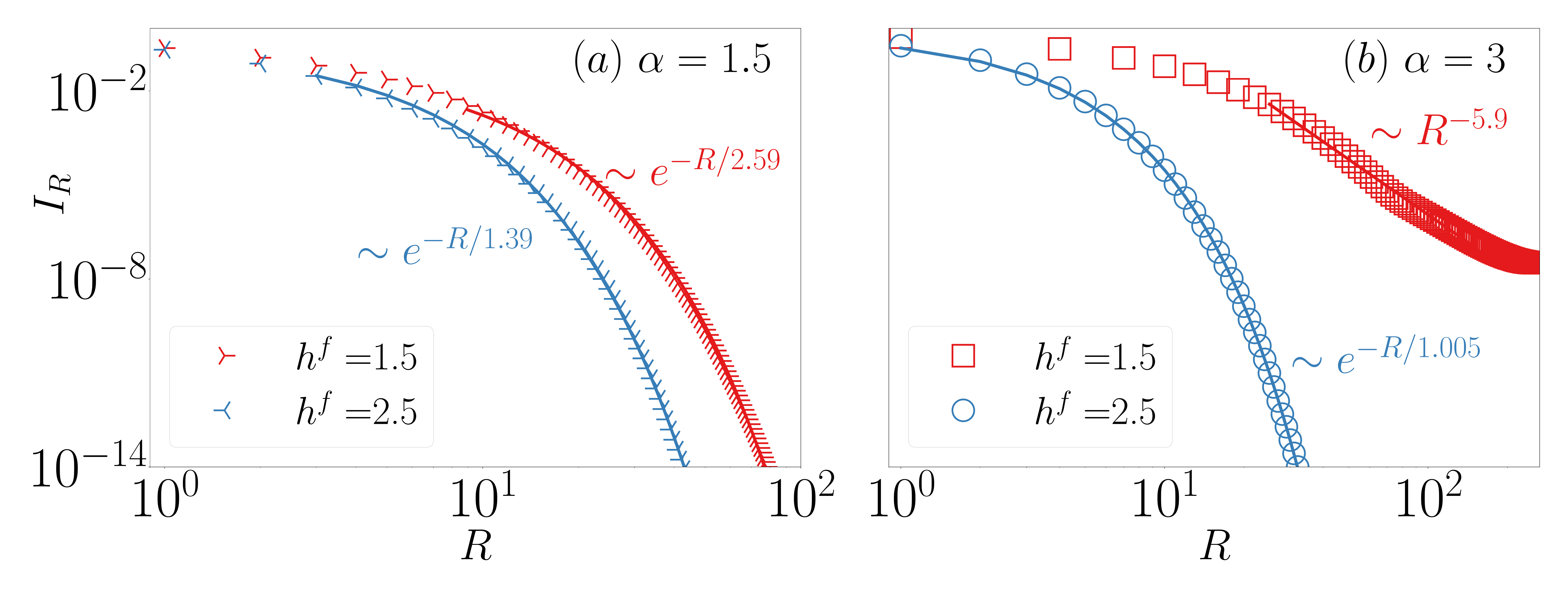}
   \caption{Steady state TC after local quench across $h$ at (a) $\alpha=1.5$ and (b) $\alpha=3$. The system is initially prepared in the ordered phase $(h^i=0.5)$ and then quenched to $h^f = \{1.5,2.5\}$.  $h^f=1.5$ corresponds to same phase quench while $h^f=2.5$ is a quench to the disordered phase across EQPT. In the quasi-local regime ($\alpha=1.5$), both the quenches exhibit exponentially decaying steady state TC. However, in the local regime ($\alpha=3$), the same phase quench has an algebraic decay while quench across EQPT features exponentially decaying TC. 
   }
   \label{fig:alpha_detetction}
\end{figure}

For FGC, by fixing the global phase at $\alpha=\alpha^f$ of the previous CGC, the prescription is as follows: $(1)$ The local parameter $h$ is initialized in the ordered phase at value $h^i$; $(2)$ two local quenches are then performed to $\{h^f_1,h^f_2\}$ where $h^f_1$ is in the same phase as $h^i$ while $h^f_2$ belongs to the other phase across the EQPT. In the quasi-local regime (Fig.\ \ref{fig:alpha_detetction}(a): $\alpha=1.5$), both the quenches lead to exponential scaling of steady state TC. Hence, in the quasi-local regime, quenches involving local EQPT make no difference in the falling rate of steady-state TC. In contrast, if both the $h$-quenches are performed in the local regime (Fig.\ \ref{fig:alpha_detetction}(b): $\alpha=3$), the same phase quench displays an algebraically decaying TC while the quench across EQPT  results in the exponential decay of TC. These contrasting feature  allow us to differentiate the local regime from the quasi-local regime.
\begin{eqnarray}
 I_R^{h^i\to h^f_{\{1,2\}}}  \sim \left \{
   \begin{array}{cc} 
          \text{same scaling} ~~&\implies  \quad \alpha^{f} < \alpha_{c}^{2}   \\
        \text{different scaling} ~~ &\implies \quad  \alpha^{f} > \alpha_{c}^{2}. 
   \end{array} \right.
   \label{scale_FGC}
\end{eqnarray}

\section{Conclusion}
\label{sec:conclusion}
 We proposed a universal criterion to identify dynamical transitions across different regimes which are characterized by different information propagation speed. Based on quantum information theoretic quantities,  we introduced \emph{coarse-} and \emph{fine-grained criteria} to determine the global transition points involved in the quench. We show that the behavior of steady state TC, which contains both classical and quantum correlation components, can identify whether a global quench involves any global transition point separating between non-local, quasi-local and local regimes. On one hand, our results unveil the potential of quantum information theoretic quantities as dynamical order parameters during global quenches. On the other hand, it shed lights on the effectiveness of local quenches in determining the global phases of the system. 
 In future studies, it would be interesting to see if a single criteria, which encapsulates both the information propagation speed and energy gap-closing, can be found to identify both the global, local dynamical transitions. 

 \begin{acknowledgments}
LGCL, SG, and ASD acknowledge the support from the Interdisciplinary Cyber Physical Systems (ICPS) program of the Department of Science and Technology (DST), India, Grant No.: DST/ICPS/QuST/Theme- 1/2019/23. This research was supported in part by the ’INFOSYS scholarship for senior students’.
DS is supported by the grant 'Innovate UK Commercialising Quantum Technologies' (application number: 44167). 
We acknowledge the use of \href{https://github.com/titaschanda/QIClib}{QIClib} -- a modern C++ library for general purpose quantum information processing and quantum computing (\url{https://titaschanda.github.io/QIClib}), and the cluster computing facility at the Harish-Chandra Research Institute. 
\end{acknowledgments}

\appendix


 \section{System Hamiltonian}
 \label{hamiltonian}

Let us we briefly discuss the family of Hamiltonian, used as the prototypical example to validate our criteria. The Hamiltonian of the LR extended-Ising model with transverse field reads as


%

\bea
\mathcal{H}
=
\sum_{n=1}^N
 \left[\frac{h'}{ 2}\sigma^z_n +
 \sum_{R=1}^{N/2} 
J'_R(\alpha)~{\cal S}_{n,R} \right].
\label{Hamil_fLR}
\eea
Here \(\sigma^\alpha_n\) (\(\alpha =x, y, z\)) are Pauli matrices at site $n$. We assume periodic boundary condition, $\sigma_{n+N}=\sigma_n$. The string operator ${\cal S}_{n,R}$ is a global operator covering  $R$ consecutive sites from $n$ to $n+R$ and has the form $${\cal S}_{n,R}=\sigma^x_n\prod_{i=n+1}^{n+R-1}\sigma^z_{i}\sigma^x_{n+R}.$$ The local and global parameters are $h'$ and $\alpha$ respectively. For this Hamiltonian,  $h^{'}$ can be understood as the local transverse magnetic field, uniform at every site while the global parameter $\alpha$ is embedded into the exchange interaction strength ${J'}_R(\alpha)$ that explicitly depends on the size of the global operator. 
We recast the system parameters as \(h'/J = h\) and \(J'_R(\alpha)/J = J_R(\alpha)\) to make them relative and dimensionless.
We further take  $J_R(\alpha)$ to be a algebraically decaying function with $R$ as 
$$J_R(\alpha)=\frac{1}{A} \frac{1}{R^{\alpha}}.$$  
The exponent $\alpha$ is positive and dictates the fall-off rate of the interaction strength with  R.  The normalization constant $A = \sum_{R=1}^{{N/2}} R^{-\alpha}$ fixes one of the the critical point at $h_c=2$ for any given $\alpha$. Although for any $\alpha<1$, $A$ is divergent in the thermodynamic limit $N\to\infty$, but in this work, we deal with finite size systems and, therefore, the normalization is still possible. For a finite size system, $A$ evaluates to be the generalized harmonic number $H_{N/2}^{(\alpha)}$, where $N$ is the size of the system while in the thermodynamic limit, this becomes the Riemann zeta function $\zeta(\alpha)$. 

 \section{Diagonalization}
\label{diagonalization}

Due to the existence of the pairwise interaction terms in the long-range  Hamiltonian, these paradigmatic Hamiltonian can be mapped to quadratic free-fermionic models which can be solved analytically. By restricting to the plus-one-parity subspace of the Hilbert space, 
we note that $H$ commutes with the parity operator, $P = \prod_{n=1}^N\sigma_n^z$. By applying the Jordan-Wigner transformation, given by

\bea
&&
\sigma^x_n~=~
 -
 \left( c_n + c_n^\dagger \right)
 \prod_{m<n}(1-2 c^\dagger_m c_m)~,\\
 \label{sigma_x}
&&
\sigma^y_n~=~
 i
 \left( c_n - c_n^\dagger \right)
 \prod_{m<n}(1-2 c^\dagger_m c_m)~, \\
 \label{sigma_y}
&&
\sigma^z_n~=~1~-~2 c^\dagger_n  c_n~, 
\label{JordanWigner}
\eea
where fermionic operators $c_n$ satisfy
$\left\{c_m,c_n^\dagger\right\}=\delta_{mn}$ and 
$\left\{ c_m, c_n \right\}=\left\{c_m^\dagger,c_n^\dagger \right\}=0$,
the Hamiltonian $H$  becomes 
\be
 H~=~P^+~H^+~P^+~+~P^-~H^-~P^-~,
\label{Hc}
\ee
where
$
P^{\pm}=\frac12\left[1\pm P\right]
$
are projectors on subspaces with even ($+$) and odd ($-$) parity 
and  
$
H^{\pm}
$
are corresponding reduced Hamiltonian. To constrain ourselves in the positive parity subspace, the fermionic operators of $H^{+}$ need to satisfy $c_{N+1}=-c_1$, the anti-periodic boundary condition.

After Jordan Wigner transformation, the Hamiltonian of the positive parity subspace reads as
\bea
H^+ &=& \sum_n \frac h2\left(1-2c_n^\dag c_n\right) \nb \\
&&+ \sum_{n,R} J_R\left[ (c_n^\dag c_{n+R}-c_nc_{n+R}^\dag) + (c_n^\dag c_{n+R}^\dag-c_nc_{n+R})\right] \nb. \\
\label{HcLR}
\eea
%
%
In the thermodynamic limit, the translationally invariant $H^+$ is diagonalized by a Fourier transformation followed by a Bogoliubov transformation.
The Fourier transformation compatible with the anti-periodic boundary condition is given by  
\be
c_n~=~ 
\frac{e^{-i\pi/4}}{\sqrt{N}}
\sum_k c_k e^{ikn}~,
\label{Fourier}
\ee
where the pseudomomentum takes half-integer values as
\be
k = \frac{(2m - 1)\pi}{N}, \text{where } m = 1, 2, \ldots \frac{N}{2}.
\label{halfinteger}
\ee
Applying the Fourier transformation, one can rewrite Eq. 
 (\ref{HcLR}) as
\bea
H^+ &=&
2\sum_{k>0} 
\left(\frac{h}{2} -  \text{Re}(\tilde{J_k})\right) \left(c_k^\dag c_k+c_{-k}^\dag c_{-k}\right) \nb \\
&&+  ~\text{Im}(\tilde{J_k})\left(c_{k}^\dag c_{-k}^\dag + c_{-k}c_{k}\right) -\frac{h}{2},
\label{HkLR}
\eea
where $\tilde{J_k} (\alpha)=\sum_{R} J_R(\alpha) e^{ikR}$ is the Fourier transformation of $J_R$. 
Therefore, we have $\tilde{J_k}(\alpha)=\frac{1}{H_{{ N/2}}^{(\alpha)}}\sum_{n=1}^{N/2}\frac{x^n}{n^\alpha}$. In the Fourier basis arranged as, $\{ |0 \rangle, c^\dagger_k|0 \rangle, c^\dagger_{-k}|0 \rangle, c^\dagger_k c^\dagger_{-k} |0 \rangle  \}$, the Hamiltonian is written as 
\begin{align}
 H^+_k =   \left[\begin{array}{cccc}
-h & 2~ \text{Im}(\tilde{J_k}) & 0 & 0 \\
2 ~\text{Im}(\tilde{J_k}) & -4 ~\text{Re}(\tilde{J_k})+h & 0 & 0 \\
0&0&-2 ~\text{Re}(\tilde{J_k})&0\\
0&0&0&-2 ~\text{Re}(\tilde{J_k})
\end{array}\right].
\label{Hk}
\end{align}

The corresponding zero-temperature state is obtained from the thermal state in the zero-temperature limit,
\begin{equation}
    \rho_0 = \lim_{\beta \to \infty} \bigotimes_{k=1}^{\frac{N}{2}} \rho_\beta^k,
\end{equation}
where the Hamiltonian is block diagonalized in the Fourier basis and the individual thermal state is defined as $ \rho_\beta^k = \frac{e^{-\beta H^+_k}}{\text{Tr}(e^{-\beta H^+_k})}$ with inverse temperature \(\beta = 1/(\kappa_B T)\), $k_B$ being the Boltzmann constant. \\

\section{Phase diagram}
\label{phases}

The dispersion relation of the system can readily be obtained upon applying the Bogoliubov tranformation on Eq.\ (\ref{HkLR})
\begin{equation}
\omega_k= 2\sqrt{(\frac{h}{2}-\text{Re}(\tilde{J_k}))^2+ \text{Im}(\tilde{J_k})^2}.
\label{omega}
\end{equation}
The transitions across the local parameter $h$ are found from the vanishing energy gap at $h_c^1=2$ and $h_c^2=-2\eta(\alpha)/\zeta(\alpha)$ \cite{Vodola2, PhysRevA.106.052425} 
corresponding to $k=0$ and $k=\pi$ respectively $\forall ~\alpha >1$ in the thermodynamics limit. Here $\eta(\cdot)$ and $\zeta(\cdot)$ are the Dirichlet eta function and Riemann zeta functions respectively. The region in between $h_c^1$ and $h_c^2$ is identified as the ordered phase (antiferromagnetic phase corresponding to the nearest-neighbor (NN) model) and the regions corresponding to $h>h_c^1$ and $h<h_c^2$ are referred as the disordered phase (paramagnetic phase of the corresponding NN model).
Although $h_c^1=2$ ceases to be a critical point for $\alpha<1$ in the thermodynamic limit as the normalization $A=\zeta(\alpha)$ becomes divergent, for finite size systems this is not the case as $A=H_{N/2}^{(\alpha)} $ is still finite. 
However, in this case the other critical point shifts to $h_c^2 = -2~\widetilde{H}_{N/2}^{(\alpha)}/H_{N/2}^{(\alpha)}$ \cite{PhysRevA.106.052425} where $\widetilde{H}_{x}^{(\cdot)}$ and $H_{x}^{(\cdot)}$ is the generalized alternating harmonic number and generalized harmonic number respectively with the sum from $R=1$ to $x$. 

The transitions across the global parameter $\alpha$ are not directly related to the closure of the spectral gap. Instead, they are associated with the speed of information propagation throughout the system \cite{SonicHorizon, PhysRevX.10.031009,*PhysRevX.13.029901,PhysRevX.10.031010,https://doi.org/10.48550/arxiv.2303.06506}. For the strongly connected LR system, maximum velocity of information transfer is set by softest quasiparticle at $k\to 0$. One can quickly check from Eq.\ ({\ref{omega}}) that the maximum group velocity of the quasiparticle, $\lim_{k\to 0} d\omega_k/dk$, is divergent when $\alpha<2$. 
Moreover, when $\alpha<1$, the energy spectrum itself is divergent in the infrared limit and the system supports instantaneous information propagation suggesting the breakdown of causality. In the intermediate regime, $1<\alpha<2$, the causal region is determined by a sub-ballistic algebraic bound.  
In contrast to the standard Lieb-Robinson bound applicable to short-range models, the LR system even in regime $\alpha>2$ adheres to a broader Lieb-Robinson bound characterized by a generalized norm which suffices for the linear free-particle light-cone effect. 
In summary, our system have three distinct regimes depending on the global parameter $\alpha$, namely (a) non-local regime ($\alpha < 1$), (b) intermediate quasi-local regime ($1< \alpha < 2$), and (c) local regime ($\alpha>2$). 

 \section{Correlation functions}
\label{pfaff_corre}

The two-point correlations are the fundamental constituents of our study. In the context of the ground state or the thermal state, the two-site correlation function between sites separated by a distance of $R$ can usually be expressed as a determinant of an $R \times R$ Toeplitz matrix.
However, as we are interested in the correlations of the time-evolved state which are, in general, not an eigenstate of the Hamiltonian, the Toeplitz determinants are not good enough. We need to deal directly with the Pfaffians in order to evaluate the correlations as a function of time as well as distance. We first compute matrix elements of operators at out-of-equilibrium in the Fourier basis and then evaluate the expectation value of the same on the time-evolved state $\rho(t) = e^{-iH^+t} \rho_0 e^{iH^+t}$ with $\rho_0$ being the initial state.

Using the Pfaffian formalism, we first write the spin correlation functions as 




\begin{widetext}
    \begin{eqnarray}
& \left .
\begin{array}{clllllcll} 
C^{lm}_{i,i+R} = \langle \sigma_i^l \sigma_{i+R}^m \rangle= {c(l,m)} {\rm pf}\; \left | \right . \mathcal{I}^{lm}_{1,2} & \dots  & \mathcal{I}^{lm}_{1,R-1}  &\mathcal{J}^{lm}_{1} & \mathcal{F}^{lm}_{1} 
& \mathcal{G}^{lm}_{1,2} &\phantom{c} .  &  \dots & \mathcal{G}^{lm}_{1,r}  \\
           & \dots  & \dots &\dots &  \dots & \dots  & \phantom{c} . & \dots & \dots \\
           &        & \mathcal{I}^{lm}_{R-2,R-1}  & \mathcal{J}^{lm}_{R-2}         & \mathcal{F}^{lm}_{R-2}         & \mathcal{G}^{lm}_{R-2,2} & \phantom{c} .  & \dots & \mathcal{G}^{lm}_{R-2,R}         \\
           &        &   & \mathcal{J}^{lm}_{R-1}          & \mathcal{F}^{lm}_{R-1} &  \mathcal{G}^{lm}_{R-1,2} & \phantom{c} .& \dots &  \mathcal{G}^{lm}_{R-1,R}  \\
           &        &   &           & \mathcal{E}^{lm} &  \mathcal{D}^{lm}_{2} & \phantom{c} . & \dots &  \mathcal{D}^{lm}_{R}  \\
           &        &   &           &             & \mathcal{K}^{lm}_{2} & \phantom{c} .&\dots &  \mathcal{K}^{lm}_{R} \\  
           &        &   &           &             &                       & \mathcal{H}^{lm}_{2,3}  &\dots &  \mathcal{H}^{lm}_{2,R} \\  
           &        &   &           &             &                     & &\dots & \dots         \\  
           &        &   &           &             &                  &   &  & \mathcal{H}^{lm}_{R-1,R}  

\end{array}
\right |,&
\label{pfaffian}
\end{eqnarray}

\end{widetext}

where $c(x,x)=c(y,y)=(-1)^{R(R+1)/2}$, 
\begin{eqnarray}
 \mathcal{I}^{xx}_{\mu,\nu}&&=  \langle A_{l+\mu}(t) A_{l+\nu}(t)\rangle, \nonumber \\
\mathcal{J}^{xx}_{\mu},  &&=\mathcal{I}^{xx}_{\mu,R}, \nonumber \\ 
\mathcal{H}^{xx}_{\mu,\nu}&&= \langle B_{l+\mu-1}(t) B_{l+\nu-1}(t)\rangle, \nonumber \\
\mathcal{K}^{xx}_{\nu}&&= \mathcal{H}^{xx}_{1,\nu}, \\ \nonumber 
\mathcal{G}^{xx}_{\mu,\nu}&&= \langle A_{l+\mu}(t) B_{l+\nu-1}(t)\rangle, \label{xx}\\ 
\mathcal{F}^{xx}_{\mu}&&=\mathcal{G}^{xx}_{\mu,1},  \nonumber \\
\mathcal{E}^{xx}&&=\mathcal{G}^{xx}_{R,1},  \nonumber \\
\mathcal{D}^{xx}_{\nu}&&=\mathcal{G}^{xx}_{R,\nu},  \nonumber 
\end{eqnarray}
\begin{eqnarray}
 \mathcal{I}^{yy}_{\mu,\nu}&&=  \langle A_{l+\mu-1}(t) A_{l+\nu-1}(t)\rangle, \nonumber \\
\mathcal{J}^{yy}_{\mu}  &&=\mathcal{I}^{yy}_{\mu,R}, \nonumber \\ 
\mathcal{H}^{yy}_{\mu,\nu}&&= \langle B_{l+\mu}(t) B_{l+\nu}(t)\rangle, \nonumber \\
\mathcal{K}^{yy}_{\nu}&&= \mathcal{H}^{yy}_{1,\nu}, \\ \nonumber 
\mathcal{G}^{yy}_{\mu,\nu}&&= \langle A_{l+\mu-1}(t) B_{l+\nu}(t)\rangle, \label{yy} \\ 
\mathcal{F}^{yy}_{\mu}&&=\mathcal{G}^{yy}_{\mu,1},  \nonumber \\
\mathcal{E}^{yy}&&=\mathcal{G}^{yy}_{R,1},  \nonumber \\
\mathcal{D}^{yy}_{\nu}&&=\mathcal{G}^{yy}_{R,\nu},  \nonumber 
\end{eqnarray}
with $s(x,y)=s(y,x)= -i (-1)^{R(R-1)/2}$, we have
\begin{eqnarray}
\mathcal{I}^{xy}_{\mu,\nu}&&=  \langle A_{l+\mu}(t) A_{l+\nu}(t)\rangle, \nonumber \\
\mathcal{G}^{xy}_{\mu,\nu}&&= \langle A_{l+\mu}(t) B_{l+\nu}(t)\rangle, \nonumber  \\
\mathcal{J}^{xy}_{\mu}  &&= \mathcal{G}^{xy}_{\mu,0}, \label{xy} \\ 
\mathcal{F}^{xy}_{\mu}&&=\mathcal{G}^{xy}_{\mu,1},  \nonumber \\
\mathcal{H}^{xy}_{\mu,\nu}&&= \langle B_{l+\mu}(t) B_{l+\nu}(t)\rangle, \nonumber \\
\mathcal{E}^{xy}&&=\mathcal{H}^{xy}_{0,1},  \nonumber \\
\mathcal{D}^{xy}_{\nu}&&= \mathcal{H}^{xy}_{0,\nu}, \nonumber \\  
\mathcal{K}^{xy}_{\nu}&&=\mathcal{H}^{xy}_{1,\nu},  \nonumber 
\end{eqnarray}
and
\begin{eqnarray}
\mathcal{I}^{yx}_{\mu,\nu}&&=  \langle A_{l+\mu-1}(t) A_{l+\nu-1}(t)\rangle, \nonumber \\
\mathcal{G}^{yx}_{\mu,\nu}&&= \langle A_{l+\mu-1}(t) B_{l+\nu-1}(t)\rangle, \nonumber  \\
\mathcal{J}^{yx}_{\mu}  &&= \mathcal{I}^{yx}_{\mu,R}, \label{yx} \\ 
\mathcal{F}^{yx}_{\mu}&&=\mathcal{I}^{yx}_{\mu,R+1},  \nonumber \\
\mathcal{E}^{yx}&&=\mathcal{I}^{yx}_{r,r+1},  \nonumber \\
\mathcal{D}^{yx}_{\nu}&&= \mathcal{G}^{yx}_{R,\nu},  \nonumber \\
\mathcal{K}^{yx}_{\nu}&&=\mathcal{G}^{yx}_{R+1,\nu}, \nonumber  \\ 
\mathcal{H}^{yx}_{\mu,\nu}&&= \langle B_{l+\mu-1}(t) B_{l+\nu-1}(t)\rangle. \nonumber 
\end{eqnarray}
Each of the elements in the Pfaffian can be constructed from the expectation values of one of the operators, $A_l A_{l+R}$, $B_l B_{l+R}$ and $A_l B_{l+R}$ with
${A}_{i}=c_{i}^{\dagger}+c_{i}, \quad {B}_{i}=c_{i}^{\dagger}-c_{i}.$
%
When recasted in the same Fourier basis as taken in Eq.\ (\ref{Hk}), the operators for the $k^\text{th}$ momentum have the matrix form
\begin{align}
 (A_l A_{l+R})_k =   \left[\begin{array}{cccc}
0 & \sin(kR) & 0 & 0 \\
-\sin(kR) & 0 & 0 & 0 \\
0&0&i\sin(kR)&0\\
0&0&0&-i\sin(kR)
\end{array}\right],
\end{align}
\begin{align}
 (B_l B_{l+R})_k =   \left[\begin{array}{cccc}
0 & \sin(kR) & 0 & 0 \\
-\sin(kR) & 0 & 0 & 0 \\
0&0&i\sin(kR)&0\\
0&0&0&-i\sin(kR)
\end{array}\right],
\end{align}

\begin{align}
 (A_l B_{l+R})_k =   \left[\begin{array}{cccc}
0 & \sin(kR) & 0 & 0 \\
\sin(kR) & -2\cos(kR) & 0 & 0 \\
0&0&-\cos(kR)&0\\
0&0&0&-\cos(kR)
\end{array}\right].
\end{align} 
The single-site transverse magnetization in the same Fourier basis is given by
\begin{eqnarray}
    \sigma_z^k = \left[\begin{array}{cccc}
1 & 0 & 0 & 0 \\
0 & -1 & 0 & 0 \\
0&0&0&0\\
0&0&0&0
\end{array}\right]. 
\end{eqnarray}
The corresponding expectation value of the operators for each value of $k$, say $O^k$, with respect to the time evolved thermal state is computed as $  \langle O \rangle = \sum_{k=1}^{{N/2}}\text{Tr}(\rho_\beta^k(t) O^k).$
Once we find all the single-site magnetizations $m_j^\alpha ~\forall~ \alpha=\{x,y,z\}$ at $j=i$ and $j=i+R$ and all possible two site correlation functions $C_{i,i+R}^{l,m} ~\forall~ l,m=\{x,y,z\}$  from the Pfaffians described above, we can construct the two-site reduced density matrix $\rho_{i,i+R}$ using Eq.\ (3) of the main text. 

\section{Total correlations in terms of quantum mutual information}
\label{mutual_info}
Let us now elucidate the total correlations in the time-evolved state as functions of the correlation functions. 
For the bipartite state $\rho_{i,i+R}$,   
the mutual information between two subsystems at any site $i$ and $i+R$ is defined as
\begin{equation}
    I_{R} = S(\rho_i) + S(\rho_{i+R}) - S(\rho_{i,i+R})
\end{equation}
 where $\rho_{i} = \mbox{tr}_{i+R} (\rho_{i,i+R})$, $\rho_{i+R} = \mbox{tr}_{i} (\rho_{i,i+R})$ and $S(\sigma) = - \mbox{tr}(\sigma \log_2 \sigma)$ is the von Neumann entropy of the system $\sigma$. The index $i$ is not relevant here as the system is translationally invariant. For the dynamical state of the translational invariant extended long-range Ising model, it reduces to
 \begin{align}
     I_{R} &= 2 H(\{\mu_1, \mu_2\}) - H(\{\gamma_1,\gamma_2, \gamma_3, \gamma_4 \} ),
 \end{align}
where \(H(\{p_i\}) = -\sum_i p_i \log_2 p_i\), with
 \begin{align}
     \mu_1 = \frac{1}{2}(1 +m_z(t)), ~
     \mu_2 = \frac{1}{2}(1  - m_z(t))\})
 \end{align}
 and
\begin{widetext}
     \begin{align}
         \gamma_1 &= \frac{1}{2}\left(1-C^{ z z}_R-\sqrt{ (C ^{xx}_R)^2+(C^{ x y}_R)^2-2 C ^{x y}_R C^{ y x}_R+ (C^{ y x}_R)^2+2C_R^{ x x} C_R^{ y y}+(C_R^{ y y})^2}\right), \nonumber\\
         \gamma_2 &= \frac{1}{2}\left(1- C_R^{ z z}+\sqrt{ (C_R^{xx})^2+ (C_R^{ x y})^2-2 C_R^{x y} C_R^{ y x}+ (C_R^{ y x})^2+2 C_R^{ x x} C_R^{ y y}+ (C_R^{ y y})^2}\right), \nonumber\\
         \gamma_3&=\frac{1}{2}\left(1+C_R^{zz}-\sqrt{(C_R^{xx})^2+ (C_R^{ x y})^2+2 C_R^{x y} C_R^{ y x}+ (C_R^{ y x})^2-2 C_R^{ x x} C_R^{ y y}+(C_R^{ y y})^2+(m^z)^2}\right), \nonumber \\
         \gamma_4&=\frac{1}{2}\left(1+C_R^{zz}+\sqrt{(C_R^{xx})^2+(C_R^{ x y})^2+2C_R^{x y} C_R^{ y x}+ (C_R^{ y x})^2-2 C_R^{ x x} C_R^{ y y}+(C_R^{ y y})^2+(m^z)^2}\right).
     \end{align}
 \end{widetext}
 Here correlation functions and magnetization of the time-evolved states can be found in the preceding sections with the help of Pfaffian.

\section{Calculation of rate function}
\label{rate_function}

Rate function is probably the most studied measure for dynamical quantum phase transition and is often used to verify whether the dynamical state still carries the signature of equilibrium quantum phase transition. Here we give a brief description of the calculation of the same. Suppose we start with the initial state $|\Psi^{in}\rangle$ which is the ground state of the Hamiltonian, $H^{in} \equiv H(\{\lambda_{0}\})$, where $\{\lambda_{0}\}$ denotes the set of the parameters in the Hamiltonian at time $t=0$. For our model, the set $\{\lambda\}$ contains a magnetic field and range of interaction as described in Sec.\ \ref{hamiltonian}. At time $t > 0$, if we perform a sudden quench by changing the values of $\{\lambda_{0}\}$ to some $\{\lambda_{1}\}$ so that the Hamiltonian is transformed from $H^{in}$ to $H^{f} \equiv H(\{\lambda_{1}\})$, the system starts evolving with time according to $H^{f}$. After time $t$, the time evolved state is given by $|\Psi(t)\rangle = e^{-iH^{f}t}|\Psi^{in}\rangle$.

To probe an equilibrium transition, say, at $\lambda_{x}$, if the initial state is prepared at $\lambda < \lambda_{x}$ and quench it with $\lambda > \lambda_{x}$, one might expect that properties of the evolved state can behave differently than the scenario when the initial state and the evolution operator both belong to the same phase, thereby recognizing the transition at $\lambda_{x}$. 
To evaluate the same, we first calculate the overlap between the initial and the time-evolved state, also known as Loschmidt amplitude, 
\begin{equation}
   \mathcal{A}(t)  = \langle \Psi^{in}|\Psi(t)\rangle = \langle \Psi^{in}|e^{-iH^{f}t}|\Psi^{in}\rangle.
   \label{amplitude}
\end{equation}
The  return 
 probability associated with this amplitude $\mathcal{A}(t)$, known as Loschmidt echo, 
\begin{equation}
\mathcal{L}(t) = |\mathcal{A}(t)|^{2},
\label{echo}
\end{equation}
 while the rate function $\mathcal{R}(t)$ corresponding to the echo for the $N$-site system is given by
\begin{equation}
    \mathcal{R}(t) = - \lim_{N \to \infty} \frac{1}{N} \log(\mathcal{L}(t)).
    \label{ret}
\end{equation}
In the non-equilibrium scenario, $\mathcal{R}(t)$ is analogous to the free energy density and similar to the EQPT, where the non-analyticity of free energy is the detector of phase transition, and here the non-analytic behavior of the rate function at the critical point carries the signature of quantum phase transition during evolution. 

Let us now concentrate on the Loschmidt amplitude for our considered system in Eq. (\ref{Hamil_fLR}). After the Jordan-Wigner and Fourier transformation, the stationary Bogoliubov–de Gennes equations become
\begin{equation}
    \omega_k
\left(
\begin{array}{c}
U_k \\
V_k
\end{array}
\right)=
2\left[\sigma^z(\frac{h}{2}-\Re(\tilde{J_k}))
 +\sigma^x\Im(\tilde{J_k})\right]
\left(
\begin{array}{c}
U_k \\
V_k
\end{array}
\right),
\label{BdGkLR}
\end{equation}
where
\be
\omega_k=2\sqrt{(\frac{h}{2}-\Re(\tilde{J_k}))^2+ \Im(\tilde{J_k})^2},
\label{omegakLR}
\ee 
is the eigenfrequencies
with $\Re(\tilde{J_k})=\sum_{r=1}^\mathcal{Z} J_r \cos{kr} $ and $ \Im(\tilde{J_k}) = \sum_{r=1}^\mathcal{Z} J_r \sin{kr}$ and $(U_k,V_k)^T$ and $(-V_k,U_k)^T$ being the corresponding eigenvectors with eigenfrequencies $+\omega_{k}$ and $-\omega_{k}$ respectively. It can be readily verifed from Eq. (\ref{BdGkLR}) that the eigenvector corresponding to the eigenvalue $+\omega_{k}$ is of the form
\begin{equation}
    \left(
\begin{array}{c}
U_k \\
V_k
\end{array}
\right)= \frac{1}{\sqrt{\mathcal{N}}}\left(
\begin{array}{c}
\frac{h}{2} - \Re(\tilde{J_k}) + \sqrt{\omega_{k}} \\
\Im(\tilde{J_k}) 
\end{array}
\right),~~~~
\label{eigenvector}
\end{equation}
with $\mathcal{N}$ being the normalization constant. 
To calculate the rate function, we need to decompose Eq. (\ref{amplitude}) in the form
\begin{equation}
    \mathcal{A}(t) = \prod_{k = 1}^{N/2} \mathcal{A}_{k}(t) =  \prod_{k = 1}^{N/2} \langle \Psi^{in}_{k}|e^{-iH^{f}_{k}t}|\Psi^{in}_{k}\rangle.
    \label{amplimomentum}
\end{equation}
The initial state of the system for a fixed value of momentum $k$ and for a particular set $\{\lambda_{in}\}$ is given by  
\begin{equation}
    |\Psi^{in}_{k}\rangle = \left(\begin{array}{c}
U_k^{in} \\ 
V_k^{in} 
\end{array} \right).
\label{ini_state}
\end{equation}
After the sudden quench from $\{\lambda_{0}\}$ to  $\{\lambda_{1}\}$, the final Hamiltonian $H^{f}$ can be expressed as diagonal in its own eigenbasis \{$(U_k^{f}, V_k^{f})^T$,  $(-V_k^{f}, U_k^{f})^T$\} 
and, therefore, the corresponding evolution is also diagonal
\begin{equation}
    \exp(-iH^{f}_{k}t) = \begin{pmatrix}
\exp(-i\omega^{f}_{k}t) & 0 \\
0 & \exp(i\omega^{f}_{k}t) 
\end{pmatrix}.
\label{unitary}
\end{equation}
\begin{figure}[b]
\includegraphics[width=\linewidth]{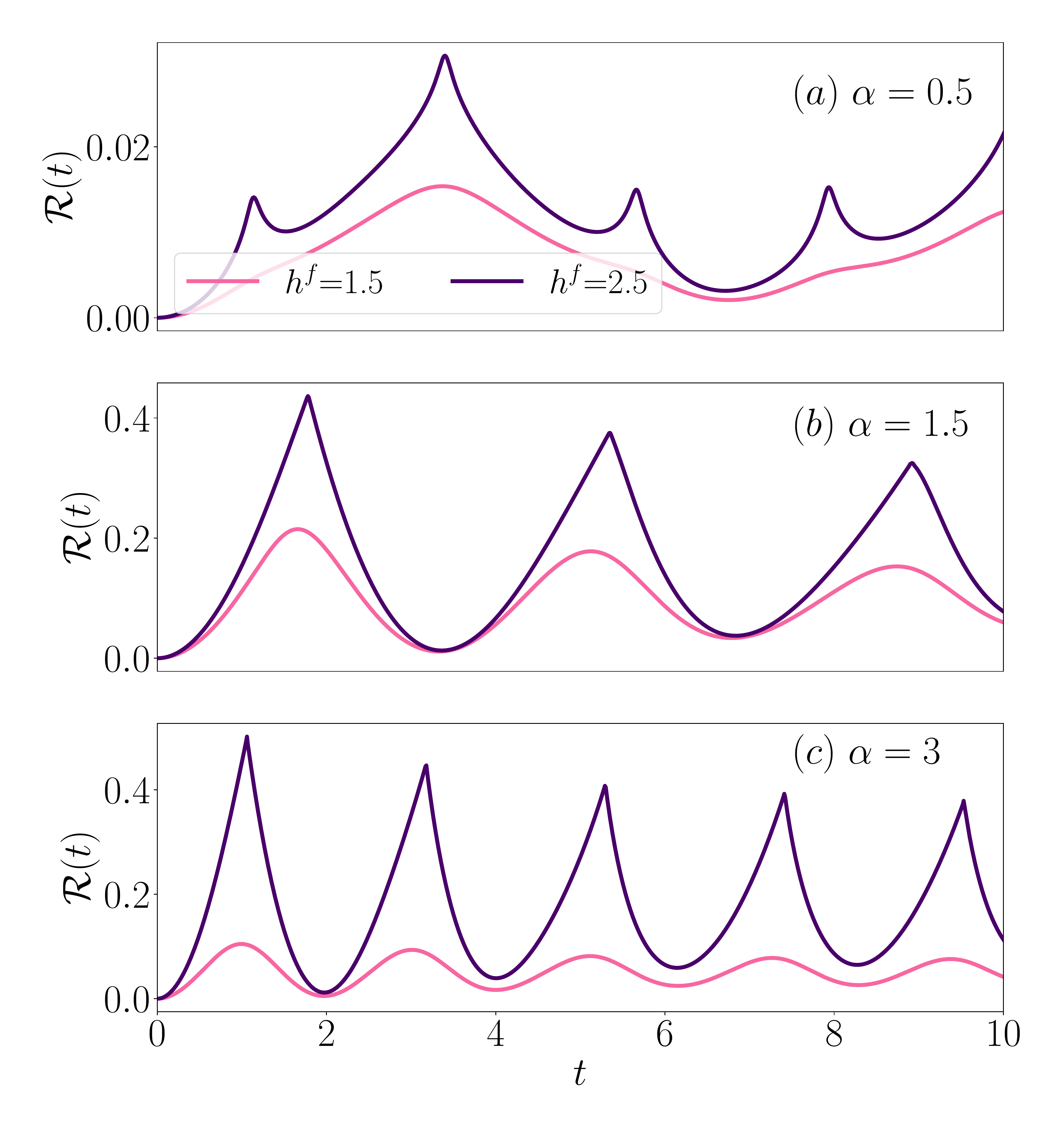}
\caption{The rate function $\mathcal{R}(t)$ against time $t$ for sudden quench across local parameter $h$ in the (a) non-local ($\alpha=0.5$), (b) quasi-local ($\alpha=1$), and (c) local ($\alpha=3$) regimes. The initial state corresponds to local parameter  $h^i = 0.5$ and is then quenched to $h^f = \{1.5, 2.5\}$. The former corresponds to the ones in the same phase with that of the initial state while the latter corresponds to different phases across the EQPT. We observe non-analyticities in the rate function $\mathcal{R}(t)$ irrespective of the parameter $\alpha$ when the local quench contains EQPT. However, when $h^f$ belongs to the same phase as $h^i$, $\mathcal{R}(t)$ is always smooth. Here $N = 512$.  }
\label{fig:h_quench}
\end{figure} 
Now, if we write the initial state $|\Psi_{k}^{in}\rangle$ in the same basis as
\begin{equation}
    |\Psi^{in}_{k}\rangle = \alpha_{k} \left(\begin{array}{c}
U_k^{f} \\ \\
V_k^{f} 
\end{array} \right) +  \beta_{k} \left(\begin{array}{c}
-V_k^{f} \\ \\
U_k^{f} 
\end{array} \right),
\label{ini_state_decomposed}
\end{equation}
the time-evolved state  simply becomes
\begin{align}
 & |\Psi^f_k(t)\rangle =  \exp(-iH^{f}_{k}t) |\Psi_{k}^{i}\rangle \nonumber \\
  &= \alpha_{k} \exp(-i\omega^{f}_{k}t ) \left(\begin{array}{c}
U_k^{f} \\ \\
V_k^{f} 
\end{array} \right) 
+ \beta_{k} \exp(i\omega^{f}_{k}t)  \left(\begin{array}{c}
-V_k^{f} \\ \\
U_k^{f} 
\end{array} \right),
\end{align}
where $|\alpha_{k}|^{2} + |\beta_{k}|^{2} = 1$ are normalization coeficients obtained from
$\alpha_{k} = \left(U_{k}^{f}, V_{k}^{f}\right) \left(
U_k^{in},
V_k^{in} 
\right)^T,$ 
and 
$\beta_{k} = \left(-V_{k}^{f}, U_{k}^{f}\right) \left(
U_k^{in}, 
V_k^{in} 
\right)^T$.
Following Eq. (\ref{amplimomentum}), the Loschmidt amplitude as a function of time turns out to be
\begin{equation}
    \mathcal{A}(t) = \prod_{k=1}^{N/2} \langle \Psi_{k}^{in} | \Psi_{k}^{f}(t) \rangle = \prod_{k=1}^{N/2} \left(|\alpha_{k}|^{2} e^{-i\omega_{k}^{f}t} + |\beta_{k}|^{2} e^{i\omega_{k}^{f}t}\right).
    \label{ampli_final_form}
 \end{equation}
 Using the same in Eq. (\ref{echo}) and (\ref{ret}), one can calculate the rate function $ \mathcal{R}(t)$. The non-analytic behavior of the same with respect to time signals the onsite of an rDQPT.

\section{The rate function for quenches across the local parameter}
\label{h_quench}

Together with global quenches discussed in the main text, let us also discuss how rate function behave for quantum quenches across the local parameter $h$ which necessarily involve a gap-closing in the energy spectrum and the critical point is characterized by universal exponent $\mu$ and $z$ which are functions of the global parameter $\alpha$.
As discussed in Appendix \ref{hamiltonian}, the LR extended Ising model possess a critical point at $h_c = 2$ which separates between the ordered and the disordered phases. To distinguish between these two phases 
we quench the local parameter  from $h^i(t = 0)$ to $h^f(t>0)$, by keeping all other parameters fixed to the respective initial values and let the system evolve. 
We show the quench dynamics via the rate function in
 Fig. \ref{fig:h_quench}.  It is clear from the Fig.\ \ref{fig:h_quench} that when the quench is within same phase and hence does not contain an EQPT ($ \{h^i, h^f\} < h_c$, -- the pink lines), the rate function is always smooth against time and hence  
 DQPT is absent in this time dynamics.
However, if the system is quenched across different phases ($h^i > h_c > h^f$, -- the purple lines) 
DQPT can be clearly seen via the non-analyticities in the rate function at certain times both in the non-local and quasi-local regimes indicating the presence of EQPT in the quench.  Irrespective of the values of the global parameter $\alpha$, the rate function can always capture the phase boundaries across the local parameter $h$. 
 Interestingly, 
steady-state TC can only capture this feature in the local regime, i.e., \ when \(\alpha>2\) by showing different scalings (see Fig.\ 5 in the main text). When $\alpha<2$, TC is always exponentially decaying and, therefore, is insufficient to signal the presence of EQPT along $h$. We use this feature in our FGC to identify the $\alpha_c^2=2$ transition point.

\bibliography{bib_v5.bib}

\end{document}